\documentclass[preprint,preprintnumbers,prb,amsmath,amssymb]{revtex4}\usepackage{fix-cm}

\usepackage{graphicx, subfigure}
\usepackage{amssymb, epstopdf}
\usepackage{natbib}
\bibpunct{(}{)}{,}{s}{}{}

\begin{document}
\bibliographystyle{unsrtnat}

\title{A theory for the compression of two dimensional strongly aggregated colloidal networks}
\author{Saikat Roy and Mahesh S Tirumkudulu\\{\it Department of Chemical Engineering}\\ {\it Indian Institute of Technology Bombay, Powai,Mumbai 400076, India}}

\begin{abstract}
The consolidation of suspended particulate matter under external forces such as pressure or gravity is of widespread interest. In this work, we derive a constitutive relation to describe the deformation of a {\it two-dimensional} strongly aggregated colloidal system by  incorporating the inter-particle colloidal forces and contact dynamics. The theory accounts for the plastic events that occur in the form of rolling/sliding during the deformation along with elastic deformation. The theory predicts a yield stress that is a function of area fraction of the colloidal packing, the coordination number, the inter-particle potential, coefficient of friction and the normal and tangential stiffness coefficients. The predicted yield stress scales linearly with area fraction for low area fractions, and diverges at random close packing. Increasing the normal stiffness coefficient or the friction coefficient increases the yield stress. For stresses greater than the yield stress, both elastic and plastic deformations contribute to the overall stress.
\end{abstract}

\maketitle

\newpage

\section*{Introduction}

The process of concentrating suspended particulate solids in liquids under the influence of an applied load such as gravitational force, centrifugal force or an applied pressure load in a filter, known as consolidation,{\cite{stickland2009whither,channell2000effects, bergstrom1992consolidation,buscall1982elastic,buscall1987consolidation, green1994exploitation,green1997yielding,landman1994solid,lange1987pressure, miller1995compressive,miller1996comparison}} is a problem of extensive practical and theoretical importance. The densification of particulate suspensions finds application in solid-liquid separation processes,\cite{landman1991dewatering,landman1988continuous} fabrication of ceramic materials{\cite{chiu1993drying,lewis2000colloidal}}  and in drying of colloidal dispersion to create particulate solids \cite{brown2002consolidation} or continuous polymer film \cite{holl2001film,tirumkudulu2005cracking}.
Consolidation of colloidal particles is influenced by a number of factors such as particle size, shape, and inter-particle potential and depends on a balance of three forces, namely, the external driving force such a gravity or centrifugal force, the viscous drag force and a particle or network stress developed as result of direct particle-particle interactions.  For colloidally stable suspensions where the particles are not in contact, the particle stress is simply the osmotic pressure of the particles whereas for flocculated or coagulated suspensions it is the elastic stress developed in the network of particles \cite{buscall1987consolidation}. In the latter case, the particles are strongly flocculated with a potential minimum being much larger than the thermal energy, $-\phi_{min}/kT \ge 20$. \cite{larson1999structure} Consequently, once the inter-particle contacts are formed the particles cannot be separated by thermal agitation. 

The behavior of irreversibly consolidating flocculated suspension under external stress, which is the focus of this work, is typically described in terms of a compressive yield stress, $P_{y}$. Here, the particle network in the flocculated dispersion spans the entire volume of the container (this occurs above the ``gelation" volume fraction) and an external compressive stress is applied on the network. The particle network along with the particles themselves deform elastically for small loads so that on removal of the load, the network and the particles recover to their respective pre-stress configuration. However, when the stress exceeds a critical value, termed as the compressive yield stress, the particles rearrange so that the network deforms permanently, and consolidates irreversibly to a new volume fraction.  At this stage, removal of load recovers only part of the total strain, which is the elastic component of the total strain, while the rest is lost due to plastic deformation. 
As the volume fraction increases, the compressive yield stress also increases since the number of contacts per particle increase and therefore the particle network is expected to resist higher loads. In addition to the particle volume fraction, the compressive yield stress is also expected to be a function of the size and shape of the particles and the inter-particle potential. Finally, the compressive yield stress diverges when the volume fraction increases to the random close packing volume fraction. At this stage, the particles cannot rearrange and all the stress goes in deforming the particles. If all the particles are purely elastic in nature, then the entire strain is recovered upon removal of load. 

A number of experimental studies{\cite{buscall1987consolidation, bergstrom1992consolidation, miller1995compressive, miller1996comparison, channell2000effects, brown2002consolidation, brown2003experimental}} have investigated the consolidation process of colloidal suspension under different process conditions such as batch sedimentation, pressure filtration and drying of aggregated suspensions. While the systems are diverse from cement pastes to polystyrene latex,  the dependence of the compressive yield stress on the volume fraction is observed to be qualitatively similar, in that a compressive yield stress was observed above the gelation volume fraction and the yield stress increased with increasing volume fraction diverging at some maximum volume fraction. In between the two volume fractions, a power-law dependence of the compressive stress with volume fraction was observed \cite{channell2000effects}.  While the observed behavior has been modeled using semi-empirical or scaling approaches,\cite{channell2000effects, brown2002consolidation} a quantitative micromechanical theory that accounts for the micro-structure, the inter-particle potential, and size and shape of the particles is missing. 

More recently, computer simulation of the consolidation of strongly aggregated {\it two dimensional} colloidal gels with fractal network under uniaxial compression was carried out and the compressive yield stress determined at varying packing fractions\cite{seto2013compressive}. They observed three distinct stages of compression, namely, the elastic-dominant regime where the work by compression is stored in the contact bonds between the particles and is purely elastic in nature, single-mode plastic regime where compression breaks large and weak particle networks into small robust structures via re-arrangement of particle positions primarily involving rolling of particles over their neighbors, and finally the multimode plastic regime where further compaction of the network occurs due to rolling and sliding of particle contacts accompanied by deformation of the particles themselves.


The goal of this work is to derive a constitutive relation for the deformation of a strongly flocculated network of colloidal particles while accounting for processes at the particle level in terms of both elastic strain in the particles and the plastic strain due to particle re-arrangement.  In doing so, we borrow heavily from the solid mechanics literature where constitutive relations for the deformation of the dense granular networks have been studied extensively.\cite{walton1978oblique,walton1987effective,digby1981effective,jenkins1993mean}It has been long known that when a network of non-colloidal elastic grains is jammed{\cite{torquato2000random}}, i.e.\ no grain can translate geometrically while all others remain fixed, further compaction of the structure under compressive stresses is possible either via particle rearrangement or by the means of grain deformation. Thus granular networks deform plastically or elastically during compaction of sediments with the former contributing to stress relaxation during the compaction.  As in the case of flocculated networks of colloidal particles, one of the most challenging problems is to provide better understanding of the onset of yielding in granular media. Some of the earliest studies\cite{drescher1972photoelastic,oda1985stress} on the the biaxial compaction of two dimensional granular particles showed that the evolution of the contact network in granular packings depends on at least three basic elements, namely, contact normals distribution, particle shape and the void distribution. It was observed that new contacts are generated in the direction parallel to the principal stress direction resulting in formation of column like load path in that direction. Thus the externally applied stress (at the boundary of the packing) induces anisotropy in the contact network of the granular packings and the anisotropy evolves as the deformation progresses. Further, it was observed that sliding is a major component of the microscopic deformation process when the inter-particle friction is low while rolling dominates when the friction is high. Later, simulations \cite{radjai1998bimodal} on the deformation of two dimensional granular particles have shown that during compaction, there are two types of particle networks - strong networks which carry the whole deviatoric load and weak networks which contribute to the average pressure. All contacts within a strong network are non-sliding whereas the entire dissipation due to sliding takes place in the weak network. Thus a complete description of the evolution of the stress versus strain relation would need to account for contributions from both networks. Some of these aspects of the micro-structure such as the anisotropy in contact and force distribution, and rolling and sliding of grains were later incorporated into the constitutive relation{\citep{nemat2000micromechanically,chang1990micromechanical, chang1995estimates,rothenburg1989analytical, rothenburgmicromechanical, cambou1995homogenization, jenkins1993mean}  to give a more realistic picture of the deformation process. In this respect, Jenkins and Strack\cite{jenkins1993mean} considered a random array of identical spherical particles interacting via non-central contact forces where the strain at the particle pair level was assumed to be identical to that at the macroscopic scale (affine deformation). While the normal component of the contact force was Hertzian, the tangential force was linearly elastic up to a critical value after which frictional sliding was considered. On determining the force versus displacement relation for both the normal and the tangential components, the local macroscopic stress was obtained by volume and orientational averaging of the force relation in a pair of contacting spheres in a unit cell. They considered the response of the particle packing in triaxial compression and evaluated the shear stress as a function of the shear strain while clearly distinguishing the contribution of the normal and tangential contact forces on the total stress. They further determined the evolution of the contact distribution, the volume change, and average plastic strain associated with the sliding displacement between particles in contact.

In this work, we build on the formulation of Jenkins and Strack\cite{jenkins1993mean} by incorporating the inter-particle colloidal forces and develop constitutive relation for irreversibly flocculated {\it two dimensional} colloidal systems. The theory accounts for the plastic events that occur in the form of rolling/sliding during the deformation along with elastic deformation. The theory predicts a yield stress that is a function of area fraction of the colloidal packing, the coordination number, the inter-particle potential, coefficient of friction and the normal and tangential stiffness coefficients. The predicted yield stress scales linearly with area fraction for low area fractions, and diverges at random close packing. Increasing the normal stiffness coefficient or the friction coefficient increases the yield stress. For stresses greater than the yield stress, both elastic and plastic deformations contribute to the overall stress. Thus the analysis presents a constitutive relation for the deformation of a two dimensional strongly aggregated dispersion in terms of the microscopic properties of the dispersion.

\section*{Theory}
Consider a two dimensional space filling aggregate of discs, each of diameter $D$, with an average particle density of $N/A$ in an area, $A$. Let $\alpha_{j}$ to be the unit vector from the center of an arbitrary disk to a contact point on its circumference. The rectangular cartesian components of the unit vector  $\alpha_{j}$ are (sin($\theta$), cos($\theta$)) where $\theta$ is the angle with the vertical (Fig \ref{fig1}(a)).
\begin{figure*}[t!]
\subfigure[]
{
 \includegraphics[scale=0.4, bb= 170 100 650 600, clip]{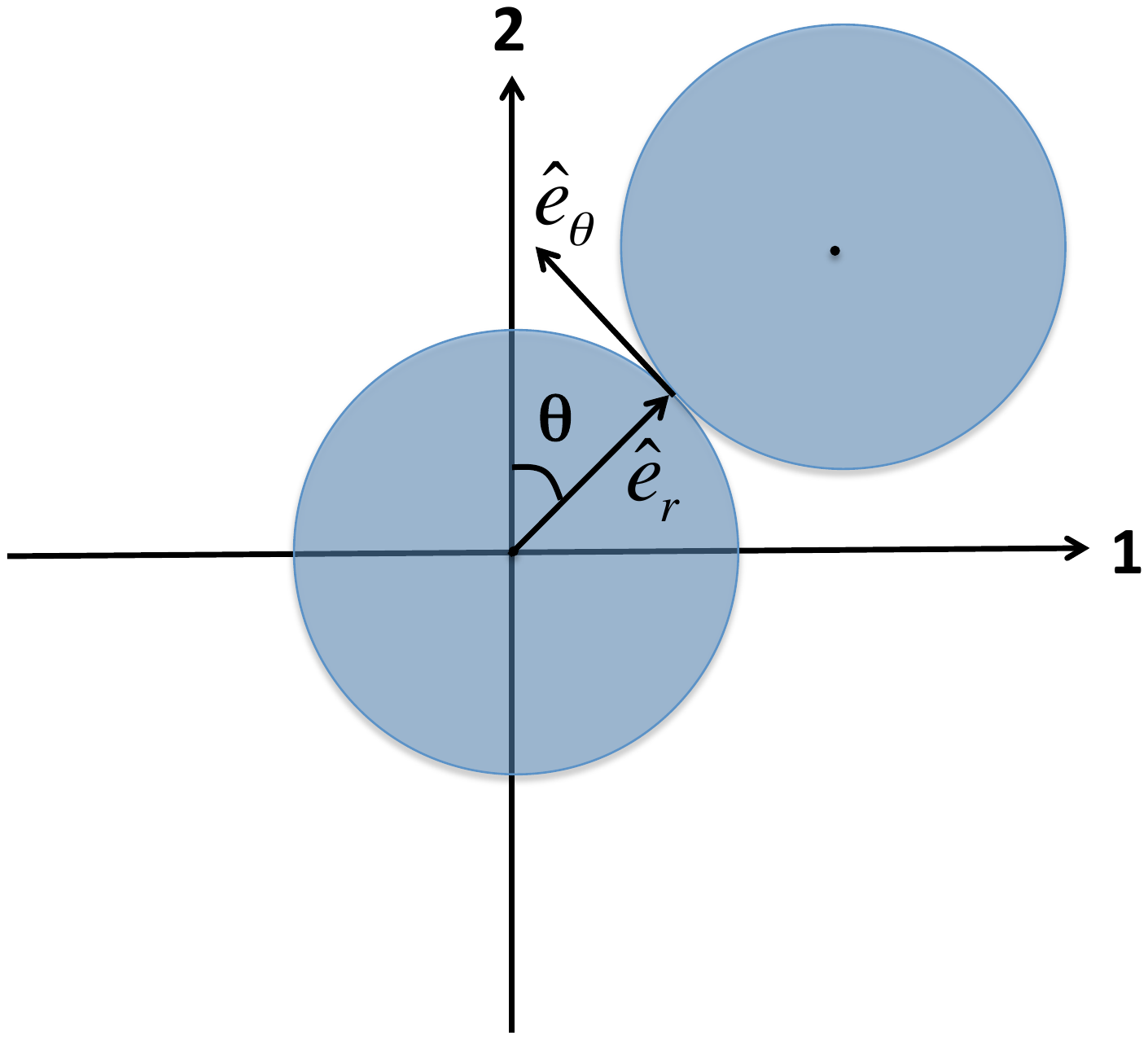}
}%
\subfigure[]
{
\includegraphics[scale=0.4, bb= 170 100 650 600, clip]{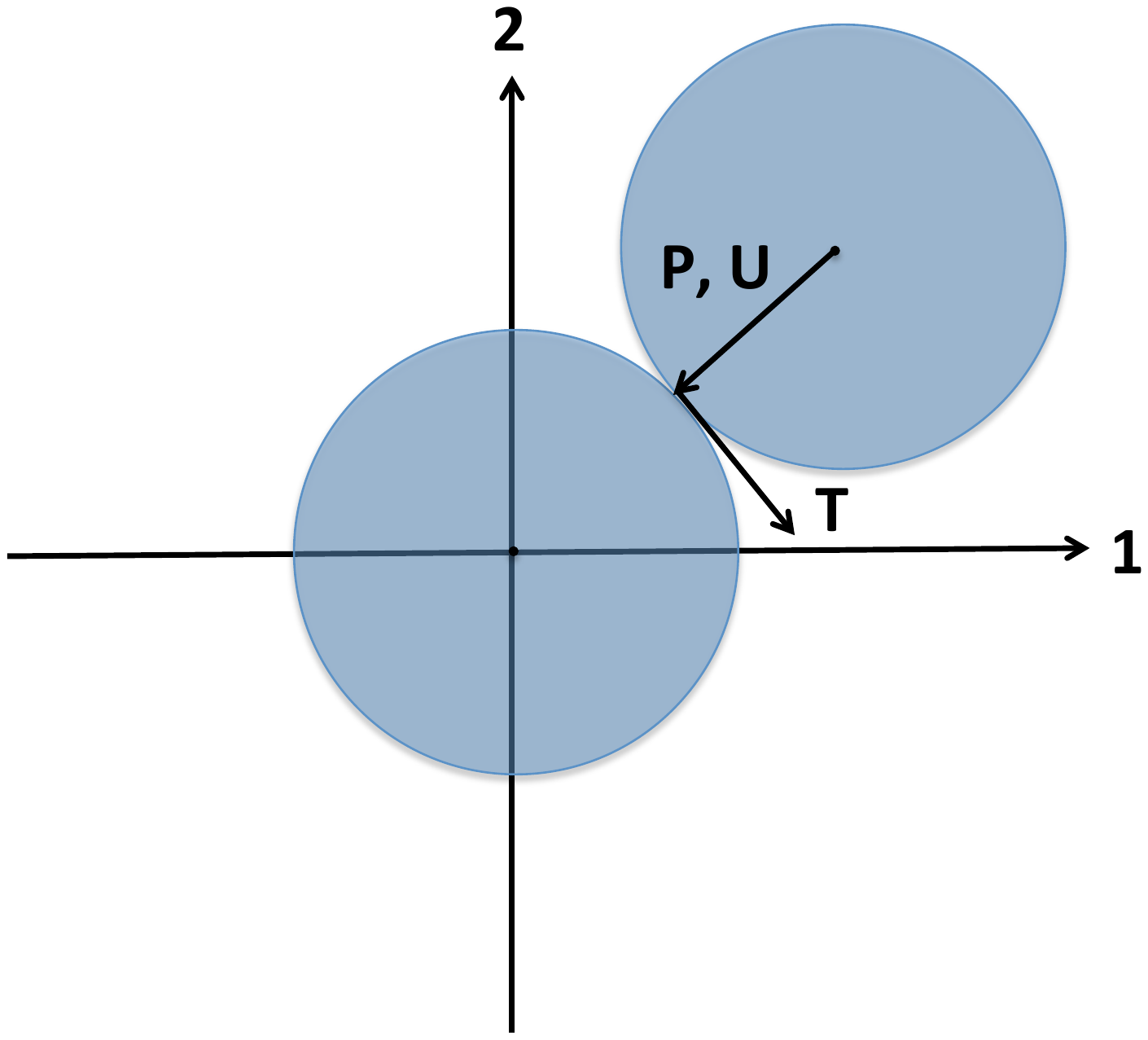}
}
\caption{(a) A two dimensional coordinate system shows the orientation of a contact point, $S$, on the circumference of the particle. The origin of coordinate system is placed at the center of a particle. Here, $\hat{e}_{r}$ and $\hat{e}_{\theta}$ represent the unit vectors in radial and theta direction. (b) Forces acting between two disks in contact for the colloidal case. Here, $P$ and $U$ are, respectively, the radial contact forces due external stress and inter-particle attraction, while $T$ is the contact force in the tangential direction.}
\label{fig1}
\end{figure*}

We assume that the deformation is affine, i.e.\ the strain at the particle pair level is the same as that of macroscopic length scale. The displacement $u_{i}$ of a contact point relative to the center of the disk is given in terms of $e_{ij}$, macroscopic strain applied at the boundary. Consequently, the externally imposed strain along with the colloidal forces between the particles result in contact forces which are depicted schematically in Figure \ref{fig1}(b). The displacement is related linearly to the strain, 
\begin{equation}
u_{i}=\frac{D}{2}e_{ij}\alpha_{j}
\end{equation}
In this work, we shall consider the uniaxial compressive strain in the `2' direction so that the strain tensor becomes,
\[e_{ij}=-\delta_{2i}\delta_{j2}e_{o}
\]
where the negative sign accounts for the compression and $e_o >0$. This assumes that the sides of the two dimensional container are rigid and does not allow expansion of the network in the `1' direction. Further, the container walls are assumed frictionless so that no shear stress is exterted by the walls on the network. The displacement of the contact point is then obtained as,
\begin{equation}
u_{i}=\frac{D}{2}e_{ij}\alpha_{j}=\left(-e_{o}cos^{2}(\theta)\hat{e}_{r}-e_{o}sin(\theta) cos(\theta)
\hat{e}_{\theta}\right)\frac{D}{2}
\end{equation}
where the magnitude of the normal component of the displacement, $\delta$, of the contact point is 
\begin{equation}
\delta=\frac{D}{2}e_{o}cos^{2}(\theta),
\label{delta}
\end{equation}
while the tangential component is
\begin{equation}
s=\frac{D}{2}e_{o}sin(\theta)cos(\theta).
\label{exps}
\end{equation}

The total contact force $F_{i}\left(\alpha\right)$ exerted by a neighboring disk at a contact point consists of three parts, namely, normal components due to inter-particle attraction, $U$ and that due to an externally imposed strain, $P$, and a tangential component, $T_i$, due to the external strain,
\begin{eqnarray}
F_{i}=-\left(P+U\right)\alpha_{i}+T_{i}.
\end{eqnarray}
 Note that the colloidal forces are assumed to be attractive and act along the line joining the centers of the particles. Further, since $T_{i}$ is perpendicular to $\alpha_{i}$, we have
\begin{equation}
T_{i}\alpha_{i}=0 
\end{equation}
Since the displacement at the particle level is due to the externally imposed strain, the force components that originate from the external strain is related to the components of the displacement. For small displacements, magnitude of the normal component of contact force is assumed to vary linearly with the normal displacement of the contact point,
\begin{equation}
P=k_{n}\delta
\end{equation}
where $k_{n}$ is the normal stiffness coefficient. Similarly the magnitude of the tangential component of the force is related to the tangential displacement,
\begin{equation}
T=k_{t}s   
\end{equation}
when $T$ is less than the critical value required for the disk to roll/slide, $T_{r}=\mu_{r}(P+U)$; the latter being the expression for Coulomb friction. Here, $k_{t}$ is the tangential stiffness coefficient and $\mu_{r}$ is the friction coefficient for rolling/sliding. Note that we do not distinguish between rolling and sliding, and instead use a common value of friction coefficient to account for particle re-arrangement during the deformation process.
Now, at the onset of rolling/sliding, the critical tangential displacement of the contact point is given by,
\begin{eqnarray}
s^{E} & = &\frac{T_{r}}{k_{t}}=\frac{\mu_{r}\left(P+U\right)}{k_{t}}=\frac{\mu_{r}\left(k_{n}\delta +U\right)}{k_{t}}\\ \nonumber
& = & \bar{\mu}_{r}\delta +\frac{\mu_{r}U}{k_{t}} \label{sE}
\end{eqnarray}
where $\bar{\mu}_{r}={\mu_{r}k_{n}}/{k_{t}}$. Thus the critical angle for the onset of rolling/sliding ($\theta_{r}$) can be obtained by substituting the expressions for $\delta$ and $s$ from (\ref{delta}) and (\ref{exps}) into (\ref{sE}),
\begin{equation}
\frac{D}{2}e_{o}\sin(\theta_{r})\cos(\theta_{r})=\bar{\mu}_{r}\frac{D}{2}e_{o}\cos^{2}(\theta_{r})+\frac{\mu_{r}U}{k_{t}}
\end{equation}
The above equation can be rearranged so as to yield a quadratic equation in $\tan(\theta)$,
\begin{equation}
\frac{2\mu_{r}U}{k_{t}De_{o}}\tan^2(\theta_{r})-\tan(\theta_{r})+\bar{\mu}_{r}+\frac{2\mu_{r}U}{k_{t}De_{o}}=0
\label{quadtan}
\end{equation}
For fixed values of parameters ($D,\bar{\mu}_{r},U,k_{t}$), the above equation will yield real solutions only above a critical value of the applied strain. In other words, no slip/roll is possible for strains below the critical value so that the entire deformation is elastic in nature. The expression for the critical strain is obtained by enforcing the condition that the roots of equation (\ref{quadtan}) be real,
\begin{equation}
e_{c}=\frac{4\mu_{r}U}{k_{t}D\left(-\bar{\mu_{r}}+\sqrt{\bar{\mu_{r}}^{2}+1} \right)}
\label{ecrit}
\end{equation}

For applied strains greater than the critical strain, equation (\ref{quadtan}) yields the critical angle, $\theta_r$, above which a neighboring particle in contact will roll/slide. Since the particles are in adhesive contact due to inter-particle attractive forces, it is assumed that there is negligible contact loss during the deformation. Therefore, any neighboring particle in contact at angles between $\theta_r$ and $\pi - \theta_r$ (and similarly for the second and third quadrant) will roll/slip while neighboring particles at all other contact points will undergo (elastic) tangential displacement. It is further assumed that the contact orientational distribution remains isotropic throughout the deformation, $E(\theta)={1}/{2\pi}$. Though this assumption does not apply to loose gels with fractal networks, we do so in the absence of an appropriate evolution equation for the structure of the particle network. We further note that presence of the side walls leads to the condition, $e_{11} = 0$, which in turns leads to biaxial compressive stress on the entire particle network. Thus, while the initial anisotropy will be strong in the `2Õ direction, it will reduce with deformation eventually leading to an isotropic distribution.  We restrict out attention to the range $0\le\theta \le {\pi}/{2}$ as configurations in the three other quadrants are the same.  Finally, the expression for the average stress tensor in a representative area or volume of the particle network can be obtained by considering traction forces acting at the boundary of the representative area \cite{love2013treatise}. For a random assembly of particles, the average stress tensor is then expressed in terms of contact orientation distribution, contact forces, contact density and branch vector joining the center of two particles,
\begin{eqnarray}
\sigma_{ij} & = & -4\frac{D}{2}\frac{Nz}{A}\int_{0}^{\pi/2}E(\theta)F_{i}\alpha_{j}d\theta\\ 
& = & -\frac{4\phi z}{\pi^{2} D}\int_{0}^{\pi/2}F_{i}\alpha_{j}d\theta 
\end{eqnarray}
where $z$ is average coordination number, and the factor of four accounts for the contribution from all the four quadrants.
The pressure in the particle network is related to the first invariant of the stress tensor,
\begin{eqnarray}
p & = & -\frac{\sigma_{ii}}{2}=\frac{2\phi z}{\pi^{2} D}\int_{0}^{\pi/2}(
-\left(P+U\right)\alpha_{i}+T_{i})\alpha_{i}d\theta\\ 
& = &\frac{2\phi z}{D\pi^{2}}\left[-\frac{k_{n}De_{o}\pi}{8}+\frac{U\pi}{2}\right]
\end{eqnarray}
Similarly, the normal stress in the `2' direction is given by,
\begin{eqnarray}
\sigma_{22}=-\frac{4\phi z}{\pi^{2}D}\int_{0}^{\pi/2}(
-\left(P+U\right)\alpha_{2}+T_{2})\alpha_{2}d\theta
\end{eqnarray}
where the first term in the integral simplifies to,
\begin{equation}
\int_{0}^{\pi/2}(\left(P+U\right)\alpha_{2}\alpha_{2})\mathrm{d}\theta=
\left(\frac{3\pi k_{n}De_{o}}{32}+\frac{U\pi}{4}\right)
\end{equation}
while the second term becomes
\begin{eqnarray}
\int_{0}^{\pi/2}T_{2}\alpha_{2}\mathrm{d}\theta=\int_{0}^{\theta_{r}}k_{t}s_{2}
\alpha_{2}\mathrm{d}\theta
-\int_{\theta_{r}}^{\pi/2}\mu_{r}\left(P+U\right)sin(\theta)\alpha_{2}\mathrm{d}\theta
\label{eqtangent}
\end{eqnarray}
Note that for the tangential part, the integration is divided into two regions, namely, ($0<\theta<\theta_{r}$) and ($\theta_{r}<\theta<\frac{\pi}{2}$) since the tangential force follows a different relationship before the start of slipping/rolling and after slipping/rolling starts. 
The rectangular components of tangential displacement are,
\begin{eqnarray}
s_{1} & = & \frac{D}{2}e_{o}\sin(\theta)\cos^{2}(\theta), \mathrm{and} \nonumber \\
s_{2} & =& -\frac{D}{2}e_{o}\sin^{2}(\theta)\cos(\theta). \nonumber
\end{eqnarray}
On substituting the above expressions in (\ref{eqtangent}) gives the following individual terms,
\begin{eqnarray}
\int_{0}^{\theta_{r}}k_{t}s_{2} \alpha_{2}\mathrm{d}\theta & = &-\frac{k_{t}D e_{o}}{16}\left[\theta_{r}-\frac{\sin(4\theta_{r})}{4}\right] \nonumber \\
\int_{\theta_{r}}^{\pi/2}\mu_{r}\left(P+U\right)\sin(\theta)\alpha_{2} \mathrm{d}\theta & =& \frac{\mu_{r}k_{n}De_{o}}{8}\cos^{4}(\theta_{r})
+\frac{\mu_{r}U}{2}\cos^2(\theta_{r})
\end{eqnarray}
The stress originating from the normal component of the contact force for all angles and the tangential component of the contact force up to the critical rolling/sliding angle is the elastic contribution to the total normal stress while that from the tangential force for angles greater than the critical rolling/sliding angle is the plastic part of the total normal stress,
\begin{eqnarray}
\left(\sigma_{22}\right)^{elastic}
& = & \frac{4\phi z}{\pi^{2}D}\left[\frac{3\pi k_{n}De_{o}}{32}+\frac{U\pi}{4}+\frac{k_{t}De_{0}}{16}\left(\theta_{r}-\frac{\sin(4\theta_{r})}{4}\right)\right], \mathrm{ and} \nonumber \\
\left(\sigma_{22}\right)^{plastic}
& = & \frac{4\phi z}{\pi^{2}D}\left[\frac{\mu_{r}k_{n}De_{o}}{8}\cos^{4}(\theta_{r})
+\frac{\mu_{r}U}{2}\cos^2(\theta_{r})\right],
\end{eqnarray}
so that the total normal stress in `2' direction contributed by both the normal and tangential component of the contact force becomes,
\begin{eqnarray}
\left(\sigma_{22}\right)^{total}  = &&  \frac{4\phi z}{\pi^{2}D}\left[\frac{3\pi k_{n}De_{o}}{32}+\frac{U\pi}{4} +\frac{k_{t}De_{o}}{16}\left(\theta_{r}-\frac{\sin(4\theta_{r})}{4}\right) \right. \nonumber \\
&& \left. + \frac{\mu_{r}k_{n}De_{o}}{8}\cos^{4}(\theta_{r}) +\frac{\mu_{r}U}{2}\cos^2(\theta_{r})\right]
\label{totalstress}
\end{eqnarray}
One of the main quantities of interest is the compressive yield stress which is the applied normal stress beyond which the packing deforms plastically. In our calculations, the compressive yield stress is equal to the the stress at the critical strain ($e_{c}$). Recall that up to the critical strain, the total stress is completely elastic as there is no rolling/sliding so that the applied stress is resisted entirely by elastic deformation of the structure. The compressive yield stress, $P_{Y}$, is obtained by substituting $\theta_{r}=\pi/2$ and $e_{o} =e_{c}$ in the expression for the total stress (\ref{totalstress}), 
\begin{equation}
P_{Y}=\frac{4\phi z}{\pi^{2}D}\left[\frac{3\pi k_{n}De_{c}}{32}+\frac{U\pi}{4}+ \frac{ \pi k_{t}De_{c}}{32}\right]
\label{ys}
\end{equation}
Note that $\left(\sigma_{22}\right)^{plastic}$ is identically equal to zero for $\theta_r=\pi/2$. We can render the yield stress non-dimensional by dividing the above expression by $U/D$,
\begin{equation}
\bar{P}_{Y}\equiv \frac{D P_Y}{U}=\frac{\phi z}{\pi}\left[1 + \frac{\bar{e}_c}{8}\left(1+ 3\frac{ k_n}{k_t}\right)\right],
\label{ysnd}
\end{equation}
where the modified critical strain is a function of only the friction coefficient and the ratio of $k_n$ and $k_t$,
\[
\bar{e}_c\equiv \frac{k_t D e_c}{U} = \frac{k_t}{k_n} \frac{4\bar{\mu}_{r}}{\left(-\bar{\mu_{r}}+\sqrt{\bar{\mu_{r}}^{2}+1} \right)}
\]
The above expressions indicate that the dimensionless yield stress is a function of only four quantities, namely, the area fraction, the coordination number, the ratio of the normal to tangential stiffness coefficient, and the friction coefficient,
\[
\bar{P}_{Y}=\bar{\sigma}_{Y}\left(\phi, z, \frac{k_t}{k_n}, \mu_{r}  \right).
\]
 For an unit area of the packing, the normal strains can be related to the area fraction before ($\phi_{in}$) and after deformation ($\phi$) via the volume conservation equation,
\[
\phi_{in} = (1 + e_{11}) (1 + e_{22}) \phi
\]
Since the side walls prevent deformation in the `2' direction ($e_{22}=0$), the area fraction is related to the strain in the normal direction, $\phi = \phi_{in}/(1 - e_o)$. In order to close the problem, the average coordination number needs to be related to the area fraction. The former should be a monotonically increasing function of $\phi$ though the exact relation would require the knowledge of the structure evolution. In the absence of such a relation, we obtained the $z$ versus $\phi$ relation by fitting the data obtained computationally for two dimensional packing by Seto et al\cite{seto2013compressive},
\begin{equation}
z=3.732 \phi^2 - 0.382\phi+ 1.984
\label{zvsphi}
\end{equation}
Substituting (\ref{zvsphi}) in (\ref{ys}) gives an explicit expression between the yield stress and the area fraction. In the current formulation, the critical strain is dependent only on the material properties (\ref{ecrit}) and is independent of the volume fraction. However, in a real situation, as the area fraction increases, neighboring particles in contact will be prevented from rolling/slipping about the particle of interest due to the presence of other  neighbors. Thus the condition for rolling/slipping derived earlier (equation (\ref{quadtan})) is strictly applicable to dilute packings where such interactions are negligible. In order to account for interactions at large area fraction, we assume the tangential friction coefficient ($\mu_{r}$) to be a function of the area fraction so that at high area fraction, larger strains are required for the particles to roll/slip. The functional form of the friction coefficient should be such that it diverges at the random close packing area fraction ($\phi_{max}$). We therefore choose a functional form similar to that used for the shear viscosity of suspension of equal sized spheres which also diverges at close packing\cite{krieger1959mechanism},
\begin{equation}
\mu_{r}(\phi)=\frac{\mu_{ro}}{\left(1-\frac{\phi}{\phi_{max}}\right)^{2}}
\label{mur}
\end{equation}
The above relation is substituted in (\ref{quadtan}) to determine the critical strain as a function of the area fraction. The latter is substituted in (\ref{ys}) to determine the variation of the yield stress as a function of the area fraction. The value of $\phi_{max}$ was set at 0.85.

\section*{Results and Discussion}

As noted above, an important outcome of the analysis is that the dimensionless yield stress depends only on the particle packing characteristics (coordination number and volume fraction) and the particle contact parameters. The absolute value of course will depend on the inter-particle force and the particle size. The former originates from a combination of electrostatic interactions and van der Waals force.  In most cases where the dispersion is strongly flocculated, the electrostatic interactions will be weak and the attractive van der Waals force will determine the inter-particle force. For the 2D case considered here, the van der Waals force can be determined by considering the interaction between two parallel cylinders of unit depth and equal diameter\cite{israelachvili2011intermolecular},
\begin{equation}
U=\frac{A_{H}}{16}\left(\frac{D}{2}\right)^{\frac{1}{2}}(\delta)^\frac{-5}{2}
\end{equation}
The value of the separation distance, $\delta$ is typically 1 to 2 nm with $A_{H}$ in the range of $1-10 \times10^{-20}$ J. For particles (or cylinders of unit length) of diameter 500 nm, this gives a force in the range of 0.01$-$0.1 N leading to a characteristic stress value ($U/D$) in the range of 20-200 kPa. The normal stiffness coefficient ($k_n$) is a measure of the rigidity of the particles and is related to the shear modulus of the particles which can vary over a very broad range, from 0.01 GPa for soft polymer particles to 100 GPa for metal oxide particles. A nominal value of $2 \times 10^{9}$ N/m is chosen for the normal stiffness co-efficient, while the ratio, $k_n/k_t$ is held fixed at $5/2$.\cite{silbert2001granular} The friction coefficient between particles, $\mu_{ro}$, could vary over a large range and we consider values of 0.1, 0.5 and 1 for the calculations. 

Figure \ref{fig3} plots the dimensionless yield stress value as a function of the area fraction. At very low $\phi$, $\mu\approx \mu_{ro}$ so that $\bar{e}_c$ is independent of $\phi$. Consequently, the dimensionless stress scales as, $\bar{P}_Y \sim z\phi$. Since in this limit, $z\rightarrow 2$, we have $\bar{P}_Y \sim \phi$. At large values of $\phi$ the functional dependency of $\mu_r$ on $\phi$ increases the power law exponent leading to a steeper rise in $\bar{P}_Y$ with $\phi$.
\begin{figure}[h]
\centering
\includegraphics[scale=0.5]{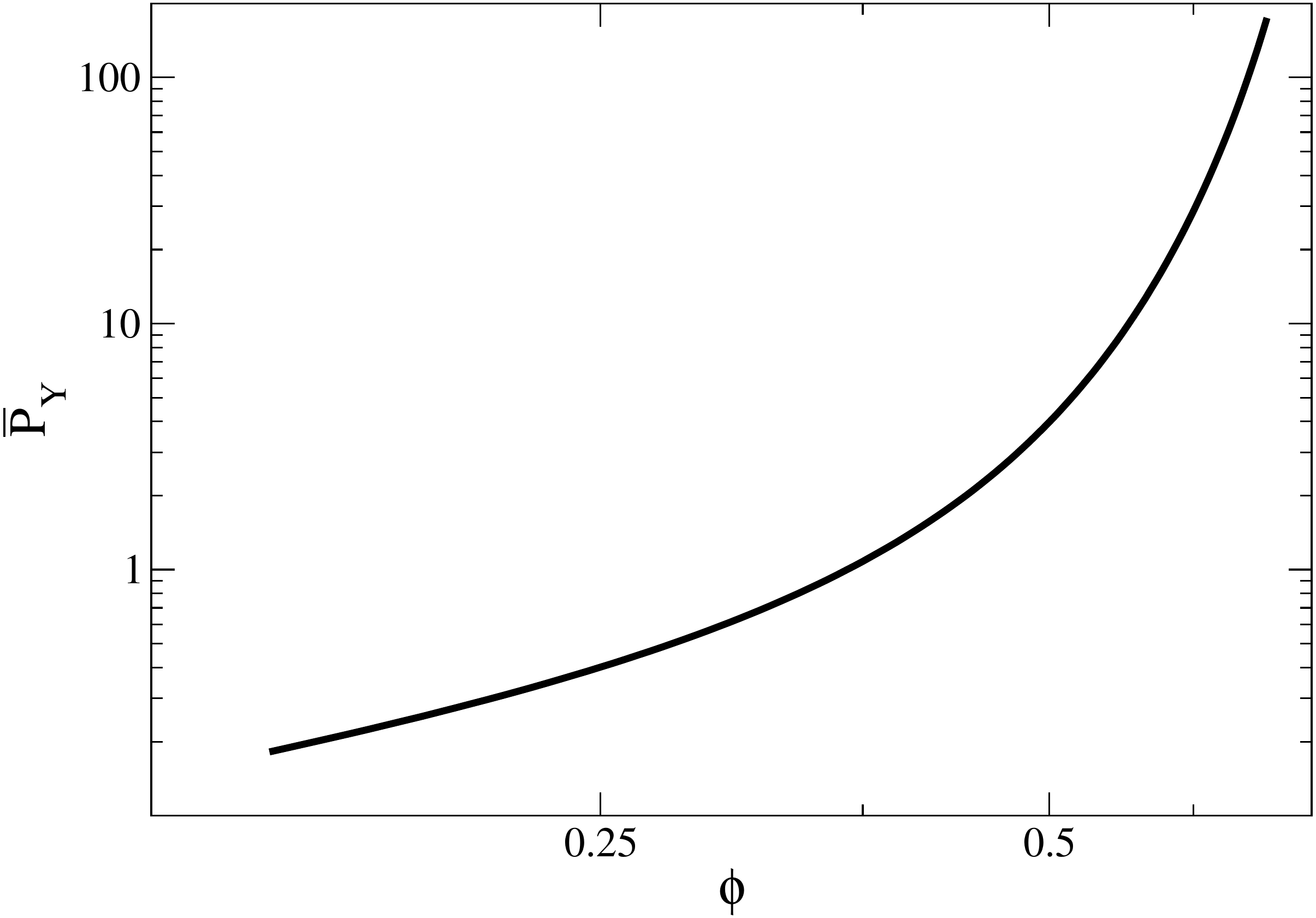}
\caption{Dimensionless yield stress vs area fraction: Here, $\mu_{ro}=0.1$ and $k_{n}/k_{t}=5/2$.}
\label{fig3}
\end{figure}
Figure \ref{fig4}(a)  plots the compressive yield stress for different $k_{n}/k_{t}$ ratios. A larger value of $k_n$ suggests that the contacts will remain elastic up to a larger strain before the tangential component induces sliding/rolling in the neighboring particles. This is reflected in the yield stress profiles where the yield stress at a fixed volume fraction increases with $k_n/k_t$. A similar effect is demonstrated in Fig \ref{fig4}(b) for three different values of $\mu_r$, where a higher value pertains to a higher resistance to sliding/rolling and therefore a higher yield stress. 
\begin{figure*}[t!]
\subfigure[]
{
 \includegraphics[height=1.9in]{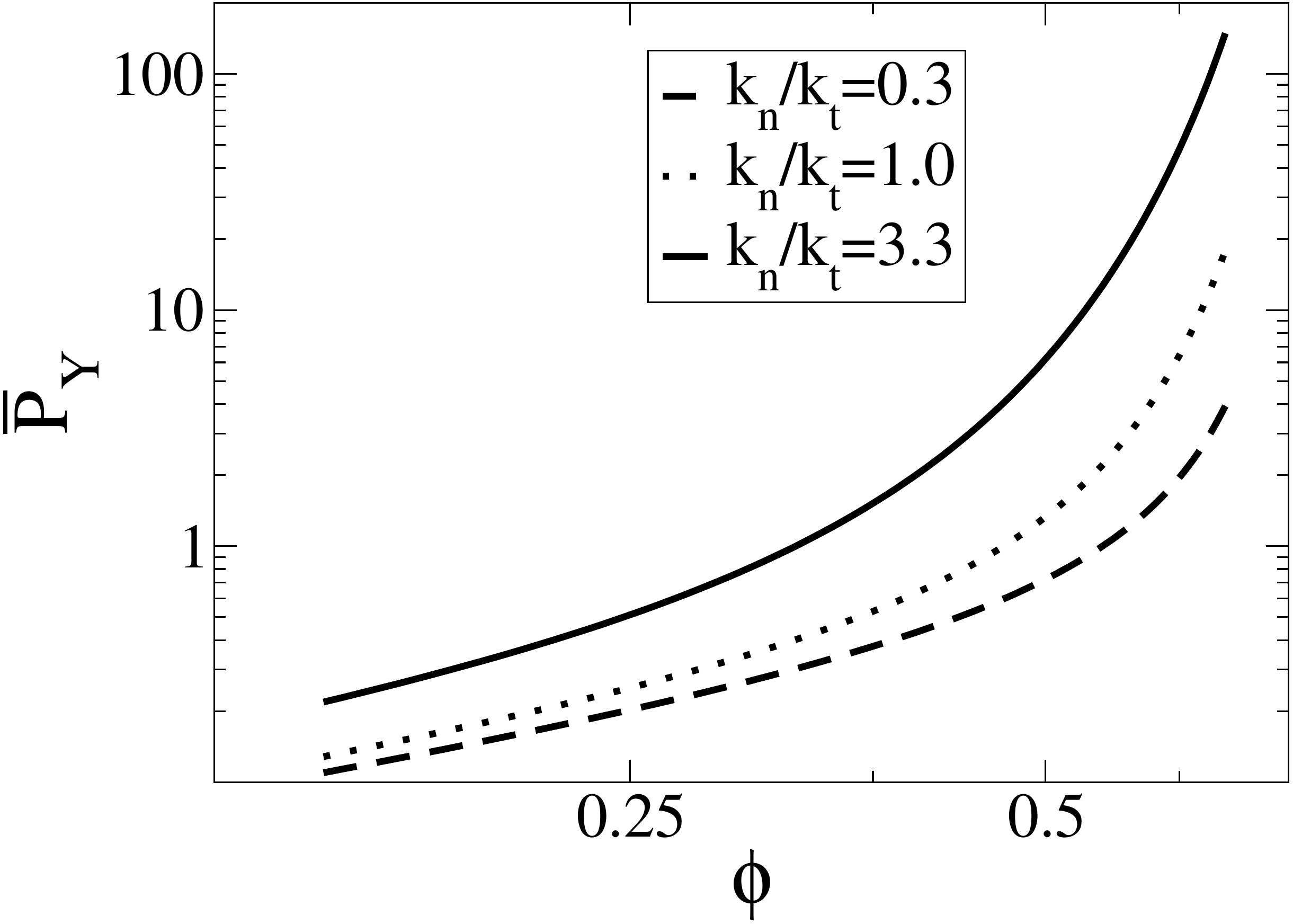}
}%
\subfigure[]
{
\includegraphics[height=1.9in]{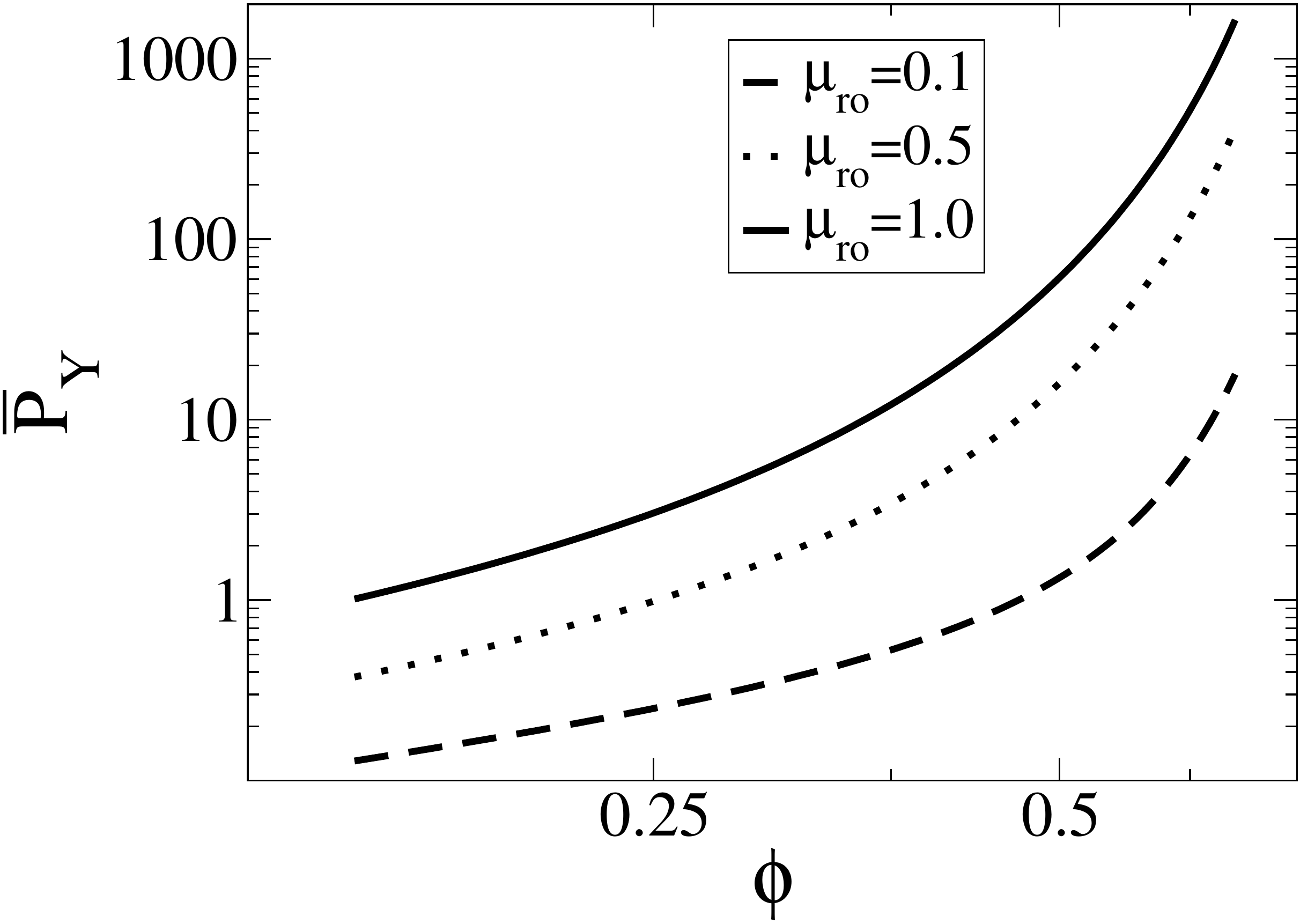}
}
\caption{Dimensionless yield stress vs area fraction for varying values of (a) $k_{n}/k_{t}$ for $\mu_{ro}=0.1$ and, (b) $\mu_{ro}$ for $k_{n}/k_{t}=1$ }
\label{fig4}
\end{figure*}

While the above plots focus on the yield stress, it is instructive to investigate how the elastic and the plastic contributions to the total stress vary as a function of axial strain for a given initial packing fraction. Figure \ref{fig5} presents the two contributions along with the total stress for an initial volume fraction of $\phi_{in}=0.30$. As expected, up to the critical strain  the total stress is solely due to elastic deformation of the network. Beyond the critical strain, particles roll/slide to yield a finite value for the plastic contribution to the total stress. With increase in strain, both the area fraction and the coordination number increase resulting in an increase in the contribution from both plastic and elastic stresses to the total stress. 
\begin{figure}[htp!]
\centering
\includegraphics[scale=0.5]{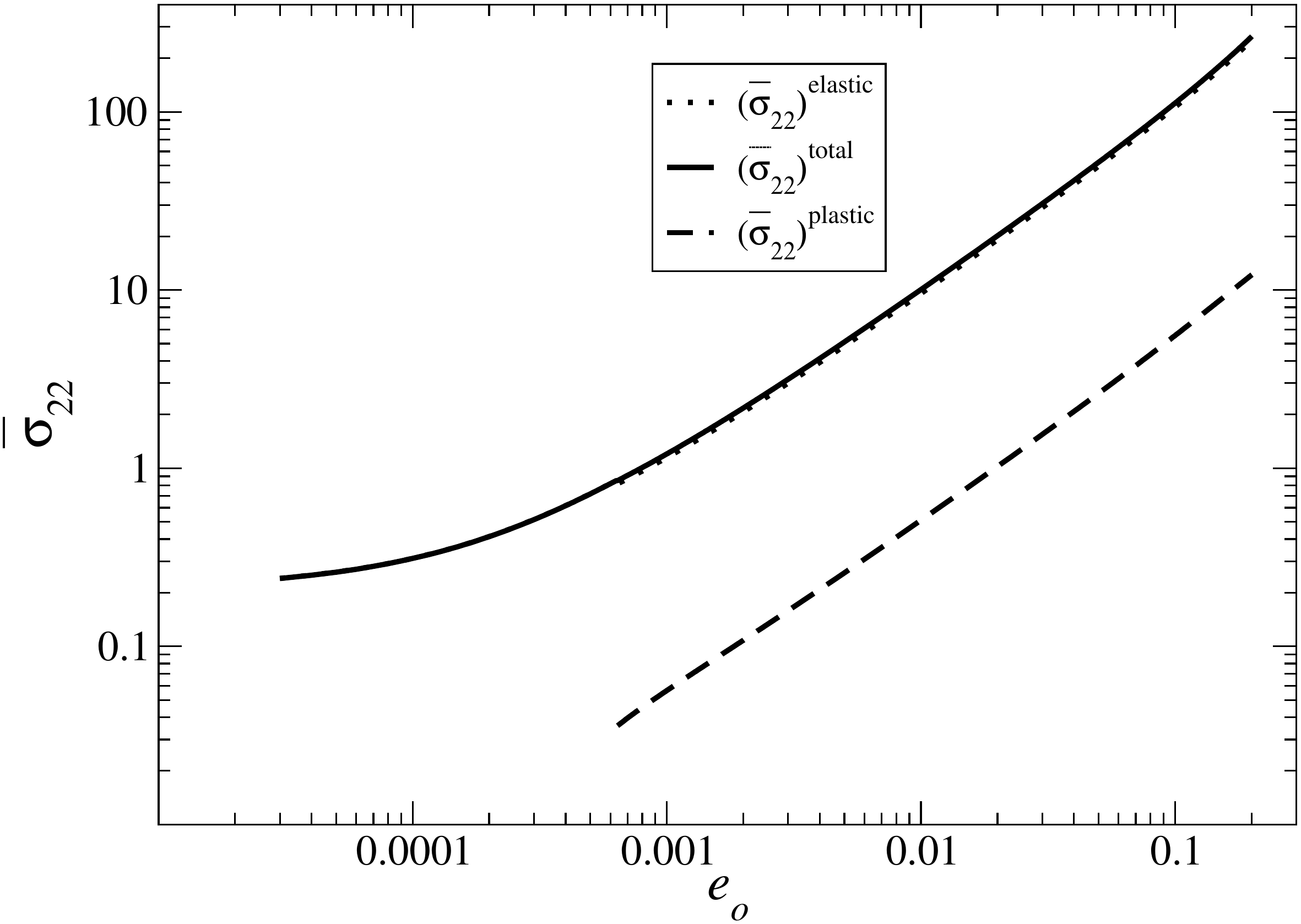}
\caption{Non-dimensionalized stress vs area fraction: elastic, plastic and total stress. Here $k_{n}/k_{t}=3.3$ and $\mu_{ro}=0.1$.}
\label{fig5}
\end{figure} 
We next compare the predicted trend for the compressive yield stress with that obtained computationally by Seto et al\cite{seto2013compressive} for two dimensional strongly aggregated colloidal gels. One of the advantages of the computational approach is that once the inter-particle interactions and the contact parameters are specified, the force balance equations can be solved for incremental strains to obtain the structure evolution of the particle network along with the yield stress. Figure \ref{fig6} presents a comparison of the predicted yield stress from our theory with that obtained computationally by Seto et al\cite{seto2013compressive}. The agreement is only qualitative and no attempt has been made to obtain a quantitative match by adjusting the parameters. The predicted trend has a slope similar to that obtained via computations at intermediate concentrations ($0.25<\phi<0.5$) from the computations.  At very low area fractions close to the gel point, the slope is determined by the dynamics of the formation of a percolating gel network and the deformation is purely elastic. As discussed previously, our analysis predicts a yield stress which varies linearly with area fractions at low area fraction while the computations yield a non-linear behavior close to the gel point.  At high area fractions ($\phi>0.5$), the predicted yield stress shows a steeper increase compared to that computed by Seto et al\cite{seto2013compressive}due to the assumed behaviour of $\mu_r$ as a function of area fraction. The plot also includes dimensionless yield stress obtained for a constant value of the friction coefficient, $\mu_r = \mu_{ro}$. Here, for intermediate and high concentrations, the power law exponent is lower and is determined by the product of the area fraction and the coordination number since the coefficient of friction is independent of the area fraction.

While the analysis presented in this paper is for a 2D aggregate, a similar analysis for the 3D case would lead to the following scaling for the compressive and shear yield stress,
\[
(P_Y, \sigma_Y) \sim \frac{z \phi U}{D^2}
\]
Experiments with well-characterized silica and polystyrene spheres show a similar scaling with particle size and potential although the scaling with volume fraction is different\cite{buscall1987consolidation}, $P_Y \sim \phi^{4.0-5.0}$ and $\sigma_Y \sim \phi^{3.0}$. Our analysis seems to suggest that the scaling of the both yield stresses with volume fraction will be the same and will critically depend on the variation of the coordination number and the friction coefficient with volume fraction.

Before closing, it is important to note the limitations of the analysis presented here. The derivation of the stress has assumed an affine deformation along with an isotropic distribution of contacts in the network. Computational studies on deformation of granular assemblies have shown that  particle slip or rolling is initiated locally and that this could result in an anisotropic distribution of both contacts and forces.  Further, the analysis assumes a functional form for both the coordination number and the friction coefficient. Despite these weaknesses, the present analysis presents a closed form solution for the yield stress of a two dimensional strongly flocculated dispersion whose trend matches qualitatively with that obtained via computations. 

\begin{figure}[htp!]
\centering
\includegraphics[scale=0.5]{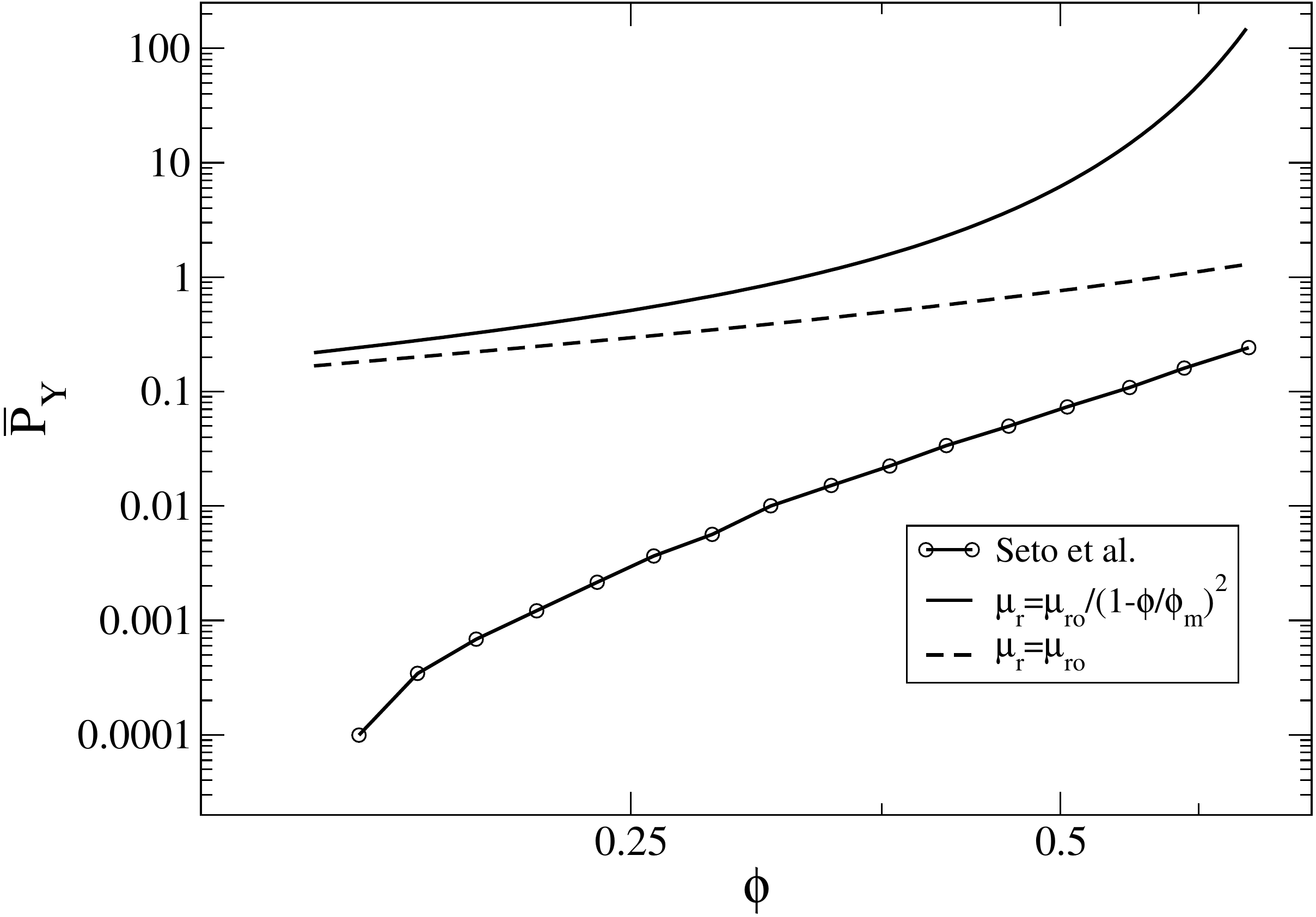}
\caption{Comparison between theoretical prediction and simulation data. \cite{seto2013compressive}Here $k_{n}/k_{t}=3.3$ and $\mu_{ro}=0.1$.}
\label{fig6}
\end{figure}

\section*{Conclusion}
In this work, we present a constitutive relation to describe the consolidation behavior of flocculated colloidal dispersion. The model accounts for the inter-particle forces, particle and contact deformation, and accounts for plastic events such as rolling/sliding during the deformation process. The particle network undergoes pure elastic deformation up to a yield stress beyond which both elastic and plastic deformation occur in the network. At very low area fractions, the compressive yield stress varies linearly with area fraction while at high area fraction the increase is steeper, with divergence at random close packing. The proposed constitutive relation depends on a few parameters that can be measured independently.  As a result, it is now possible to solve for the deformation of strongly aggregated colloidal dispersions in complex geometries. Future work will extend the analysis to three dimensional aggregated colloidal systems.

\newpage

\bibliography{ref}

\end{document}